\documentclass[a4paper]{jpconf}
\usepackage{graphicx}
\begin{document}
\title{BTZ Solutions on Codimension-2 Braneworlds}

\author{B. Cuadros-Melgar$^1$, E. Papantonopoulos$^2$, M. Tsoukalas$^2$, V. Zamarias$^2$}

\address{$^1$Departamento de F\'isica,
Universidad de Santiago de Chile, Casilla 307, Santiago, Chile}
\address{$^2$Department of Physics, National Technical
University of Athens, GR~157~73~Athens, Greece}

\ead{berthaki@gmail.com, lpapa@central.ntua.gr, minasts@central.ntua.gr, zamarias@central.ntua.gr}

\begin{abstract}
We consider five-dimensional gravity with a Gauss-Bonnet term in the bulk and 
an induced gravity term on a 2-brane of codimension-2. We show that this system
admits BTZ-like black holes on the 2-brane which are extended into the bulk 
with regular horizons.
\end{abstract}

\section{Introduction}

Codimension-2 braneworlds, i.e, a brane embedded in a bulk with two 
extra dimensions, have recently stimulated a growing insterest. 
The most attractive feature of these models is that the
vacuum energy of the brane instead of curving the brane world-volume, 
merely induces a deficit angle in the bulk  around the brane~\cite{Chen:2000at}.This property was used to solve the cosmological constant problem~\cite{6d}.
However, soon it was realized~\cite{Cline} that one can only find
nonsingular solutions if the brane stress tensor is proportional
to its induced metric. To obtain the Einstein equation on the
brane one has to introduce a cut-off (brane
thickness)~\cite{Kanno:2004nr},  losing the predictability of the
theory, or alternatively, one can add to the gravitational action 
a Gauss-Bonnet (GB) term~\cite{Bostock:2003cv} or a scalar curvature term
(induced gravity) on the brane~\cite{Papantonopoulos:2005ma}.

Time dependent cosmological solutions in this scenario are also a complex
topic. In the thin brane limit, since the brane and bulk energy momentum 
tensors are related, we cannot get the standard cosmology on
the brane~\cite{Kofinas:2005py}. Thus, one has to regularize the 
codimension-2 branes by introducing some thickness and then consider matter 
on them~\cite{regular}, arriving to a time-dependent bulk solution. Or 
alternatively, one can conceive a codimension-1 brane moving in the 
regularized static background~\cite{Papantonopoulos:2007fk}. However, the 
resulting cosmology is unrealistic having a negative Newton's constant.

Moreover, the issue of localization of a black hole on the brane and 
its extension to the bulk is not fully understood.
In codimension-1 braneworlds, the most natural generalization of the 
four-dimensional Schwarzschild metric was to consider a black string 
infinite in the fifth dimension~\cite{Chamblin:1999by}. However, although the
curvature scalars are everywhere finite, the Kretschmann scalar
diverges at the AdS horizon at infinity, which turns the above
solution into a physically unsuitable object. It has been argued that 
there exists a localized black cigar solution with a finite extension 
along the extra dimension due to a Gregory-Laflamme~\cite{glf} type of 
instability near the AdS horizon, although no analytical solution has 
been found until the present time.
Further attempts deal with the Einstein equations projected on the 
brane~\cite{SMS}, which include an unknown bulk dependent term, the 
Weyl tensor projection. 
Due to this reason the system is not closed, and some assumptions have to 
be made either in the form of the metric or in the Weyl term~\cite{BBH}.
The stability and thermodynamics of these solutions were worked out in~
\cite{ACMPM}.

A lower dimensional version of a black hole living on a (2+1)-dimensional 
braneworld was considered in~\cite{EHM}. The authors based their analysis 
on the C-metric~\cite{Kinnersley:zw} modified by a cosmological
constant term. After some coordinate transformations, the resulting 
geometry corresponds to a BTZ black hole~\cite{Banados:1992wn}
on the brane. This object can be extended as a BTZ string in a
four-dimensional anti-de Sitter (AdS) bulk, which has been cut in order
to avoid the conical singularity along the symmetry semi-axis. 
Their thermodynamical stability analysis showed that the black string 
remains a stable configuration when its transverse size is comparable to 
the four-dimensional AdS radius, being destabilized by the
Gregory-Laflamme instability above that scale, breaking up to a
BTZ black hole on a 2-brane.

In the codimension-2 scenario, a six-dimensional black hole on a 
3-brane was proposed in~\cite{Kaloper:2006ek}. 
This is a generalization of the $4D$ Aryal, Ford, 
Vilenkin~\cite{Aryal:1986sz} black hole pierced by a
cosmic string adjusted to the codimension-2 branes with a conical
structure in the bulk and deformations fitting the deficit
angle. However, it is not clear how to realize these solutions in
the thin 3-brane limit.

In this work we study black holes on an infinitely thin conical
2-brane and their extension into the five-dimensional bulk.

\section{The Setup}

We consider the following gravitational action in five dimensions
with a Gauss-Bonnet (GB) term in the bulk and an induced
three-dimensional curvature term on the brane
\begin{eqnarray}\label{AcGBIG}
S_{\rm grav}&=&\frac{M^{3}_{5}}{2}\left\{ \int d^5 x\sqrt{-
g^{(5)}}\left[ R^{(5)}
+\alpha\left( R^{(5)2}-4 R^{(5)}_{MN}R^{(5)MN}+
R^{(5)}_{MNKL}R^{(5)MNKL}\right)\right] \right.\nonumber\\
&+& \left. r^{2}_{c} \int d^3x\sqrt{- g^{(3)}}\,R^{(3)}
\right\}+\int d^5 x \mathcal{L}_{bulk}+\int d^3 x
\mathcal{L}_{brane}\,,\label{5daction}
\end{eqnarray}
where $\alpha\, (\geq0)$ is the GB coupling constant, 
$r_c=M_{3}/M_{5}^3$ is the induced gravity ``cross-over" scale, 
which marks the transition from 3D to 5D gravity, and $M_{5}$,   
$M_{3}$ are the five and three-dimensional Planck masses, respectively.

The above induced term has been
written in the particular coordinate system in which the metric is
\begin{equation} 
ds_5^2=g_{\mu\nu}(x,\rho)dx^\mu
dx^\nu+a^{2}(x,\rho)d\rho^2+b^2(x,\rho)d\theta^2~.\label{5dmetric} 
\end{equation}
Here $g_{\mu\nu}(x,0)$ is the brane metric, whereas $x^{\mu}$
denotes three  dimensions, $\mu=0,1,2$, and $\rho,\theta$ denote
the radial and angular coordinates of the two extra dimensions
(the $\rho$ direction may or may not  be compact, and the $\theta$ 
coordinate ranges form $0$ to $2\pi$).
Capital $M$,~$N$ indices will take values in the five-dimensional
space. Note that we have assumed that there exists an azimuthal
symmetry in the system, so that both the induced three-dimensional
metric and the functions $a$ and $b$ do not depend on $\theta$.

The Einstein equations resulting from the variation of the
action~(\ref{5daction}) are 
\begin{equation}
 G^{(5)N}_M + r_c^2
G^{(3)\nu}_\mu g_M^\mu g^N_\nu {\delta(\rho) \over 2 \pi b}-\alpha
H_{M}^{N} =\frac{1}{M^{3}_{5}} \left[T^{(B)N}_M+T^{(br)\nu}_\mu
g_M^\mu g^N_\nu {\delta(\rho) \over 2 \pi b}\right]~, \label{einsequat3}
\end{equation} 
where 
\begin{eqnarray}
 H_M^N&=& \left[{1
\over 2}g_M^N (R^{(5)~2}
-4R^{(5)~2}_{KL}+R^{(5)~2}_{ABKL})\right.-2R^{(5)}R^{(5)N}_{M}\nonumber\\
&&+4R^{(5)}_{MP}R^{NP}_{(5)}\phantom{{1 \over 2}}~\left.
+4R^{(5)~~~N}_{KMP}R_{(5)}^{KP} -2R^{(5)}_{MKL P}R_{(5)}^{NKL
P}\right]~. 
\label{gaussbonnet} 
\end{eqnarray} 
To obtain the braneworld
equations we expand the metric around the brane as 
\begin{equation}
b(x,\rho)=\beta(x)\rho+O(\rho^{2})~. 
\end{equation} 
At the boundary of the
internal two-dimensional space where the 2-brane is situated the
function $b$ behaves as $b^{\prime}(x,0)=\beta(x)$, where a prime
denotes derivative with respect to $\rho$. We also demand that the
space in the vicinity of the conical singularity is regular, which
imposes the supplementary conditions, $\partial_\mu \beta=0$ and
$\partial_{\rho}g_{\mu\nu}(x,0)=0$~\cite{Bostock:2003cv}.

The extrinsic curvature in the particular gauge $g_{\rho \rho}=1$
that we are considering  is given by $K_{\mu\nu}=g'_{\mu\nu}$.
 The above decomposition will be helpful in the following for finding the induced dynamics on the brane.
We will now use the fact that the second derivatives of the metric
functions contain $\delta$-function singularities at the position of
the brane. The nature of the singularity then gives the following
relations \cite{Bostock:2003cv} 
\begin{eqnarray}
{b'' \over b}&=&-(1-b'){\delta(\rho) \over b}+ {\rm non-singular~terms}~,\\
{K'_{\mu\nu} \over b}&=&K_{\mu\nu}{\delta(\rho) \over b}+ {\rm
non-singular~terms}~. 
\end{eqnarray}

From the above singularity expressions and using the Gauss-Codazzi
equations we can  match the singular parts of the Einstein
equations (\ref{einsequat3}) and get the following ``boundary"
Einstein equations
 \begin{equation} 
G^{(3)}_{\mu\nu}={1 \over M_{(5)}^3 (r_c^2+8\pi
(1-\beta)\alpha)}T^{(br)}_{\mu\nu}+{2\pi (1-\beta) \over r_c^2+8\pi
(1-\beta)\alpha}g_{\mu\nu} \label{einsteincomb3}~. 
\end{equation}
The effective three-dimensional
Planck mass and cosmological constant are simply
\begin{eqnarray}
 M^{2}_{Pl}&=&M_{(5)}^3 (r_c^2+8\pi (1-\beta)\alpha)~,\\
\Lambda_{(3)}&=&\lambda-2\pi M_{(5)}^3 (1-\beta)~, 
\end{eqnarray}
where $\lambda$ is the brane tension.  Note that the Planck mass can 
depend on the deficit angle. This is an effect of solely the bulk 
Gauss-Bonnet term.

As a result of the Gauss-Codazzi reduction, the above boundary 
Einstein equations will also contain 
terms proportional to the extrinsic curvature and additional terms coming
from the bulk GB term. However, if we allow only regular conical
singularities there is no contribution from these
terms~\cite{Bostock:2003cv}.

\section{Black String Solutions}

We assume that there is a localized (2+1) black hole on the brane described
by the following metric 
\begin{equation}
ds_{3}^{2}=\left(-n(r)^{2}dt^{2}+n(r)^{-2}dr^{2}+r^{2}d\phi^{2}\right)~,
\label{3dmetric}
\end{equation} 
where $0\leq r< \infty$ is the radial coordinate, and $\phi$ has the 
usual periodicity $(0,2\pi)$.
We will look for black string solutions of the Einstein equations~
(\ref{einsequat3}) using the five-dimensional metric~(\ref{5dmetric}) 
in the form
\begin{equation} 
ds_5^2=f^{2}(\rho)\left(-n(r)^{2}dt^{2}+n(r)^{-2}dr^{2}+r^{2}
d\phi^{2}\right)+d\rho^2+b^2(r,\rho)d\theta^2~,
\label{5smetricc}
\end{equation}
where we have chosen $a(r,\rho)=1$ without loss of generality.

The space outside the conical singularity is regular, therefore, we  
demand that the warp function $ f(\rho) $ is also regular everywhere. 
We assume that there is only a cosmological constant $\Lambda_{5}$
in the bulk. Then, the bulk Einstein equations take the form
\begin{equation} 
G^{(5)}_{MN}-\alpha
H_{MN}=-\frac{\Lambda_{5}}{M^{3}_{5}}g_{MN}~.
\end{equation} 
By combining the $(rr)$ and $(\phi \phi)$ equations we get
\begin{equation}
\left(\dot{n}^{2}+n \ddot{n}-\frac{n \dot{n}}{r}\right)\left(1-4\alpha \frac{b''}{b}\right)=0~,\label{173}
\end{equation}
while a combination of the $(\rho \rho)$ and $(\theta \theta)$ equations
gives
\begin{equation}
\left(f''-\frac{f'b'}{b}\right)\left[3-4\frac{\alpha}{f^{2}}\left(\dot{n}^{2}+n
 \ddot{n}+2\frac{n \dot{n}}{r}+3f'^{2}
 \right)\right]=0
\label{183}~,
\end{equation}
where a dot implies derivatives with respect to $r$. The solutions 
of the equations (\ref{173}) and (\ref{183}) are summarized in the 
following table~\cite{CuadrosMelgar:2007jx}.\\

\begin{table}[here]
\begin{center}
\begin{tabular}{cccccc}
\br
  $n(r)$ & $f(\rho)$ & $b(\rho)$ & $-\Lambda_5$ & Constraints \\
\mr
  BTZ & $\cosh\left(\frac{\rho}{2\,\sqrt{\alpha}}\right)$ & $\forall b(\rho)$ &
$\frac{3}{4\alpha}$ &
$L_3^2=4\,\alpha$ \\
  BTZ & $\cosh\left(\frac{\rho}{2\,\sqrt{\alpha}}\right)$ & $2\,\beta\,\sqrt{\alpha}\,\sinh\left(\frac{\rho}{2\,\sqrt{\alpha}}\right)$ &
  $\frac{3}{4\alpha}$ & - \\
  BTZ & $\cosh\left(\frac{\rho}{2\,\sqrt{\alpha}}\right)$ & $2\,\beta\,\sqrt{\alpha}\,\sinh\left(\frac{\rho}{2\,\sqrt{\alpha}}\right)$ &
  $\frac{3}{4\alpha}$ & $L_3^2=4\,\alpha$ \\
  BTZ & $\pm 1$ & $\frac{1}{\gamma}\,\sinh\left(\gamma\,\rho\right)$ & $\frac{3}{l^2}$ & $\gamma=\sqrt{-\frac{2\Lambda_5}{3+4\alpha\Lambda_5}}$ \\
  $\forall n(r)$ & $\cosh\left(\frac{\rho}{2\,\sqrt{\alpha}}\right)$ & $2\,\beta\,\sqrt{\alpha}\,\sinh\left(\frac{\rho}{2\,\sqrt{\alpha}}\right)$ &
  $\frac{3}{4\alpha}$ & -  \\
  Corrected BTZ & $\cosh\left(\frac{\rho}{2\,\sqrt{\alpha}}\right)$ &
$2\,\beta\,\sqrt{\alpha}\,\sinh\left(\frac{\rho}{2\,\sqrt{\alpha}}\right)$ &
$\frac{3}{4\alpha}$ &
$L_3^2=4\,\alpha$ \\
  Corrected BTZ & $\pm 1$ & $2\,\beta\,\sqrt{\alpha}\,\sinh\left(\frac{\rho}{2\,\sqrt{\alpha}}\right)$
  & $\frac{1}{4\alpha}$ & $\Lambda_5=-\frac{1}{4\alpha}=-\frac{3}{L_3^2}$ \\
\br
\end{tabular}
\end{center}
\caption{BTZ String-Like Solutions in Five-Dimensional Braneworlds
of Codimension-2}\label{table1}
\end{table}
where $L_3$ is the length of three-dimensional AdS space. The BTZ solution 
is given by \cite{Banados:1992wn}
\begin{equation}
n^{2}(r)=-M+\frac{r^{2}}{L_3^{2}}~.\label{btz} 
\end{equation}
This black hole for positive mass has a horizon at $r=L_3\sqrt{M}$, and the
radius of curvature of the $AdS_3$ space $L_3=(-\Lambda_3)^{-1/2}$
provides the length scale necessary to define this horizon. For
the mass $-1<M<0$, which is dimensionless, the BTZ black hole has a
naked conical singularity while for $M=-1$ the vacuum $AdS_3$
space is recovered.

Moreover, the ``corrected'' version of the BTZ black hole is found to be
\begin{equation}
n^{2}(r)=-M+\frac{r^{2}}{L_3^{2}}-\frac{\zeta}{r}~.\label{btzc} 
\end{equation}
This black hole corresponds to the BTZ black hole conformally coupled to 
a massless scalar field~\cite{Zanelli1996}. The scalar field does not 
introduce an independent conserved charge, for this reason it does not 
show up in the form of the metric. It merely modifies the energy-momentum
tensor of the three-dimensional Einstein equations as we will see in the 
next section.

Although $n(r)$ remains undetermined in one of our solutions, it exists 
a connecting relation between this function and the matter on the brane
$T_{1}^{1}=T_{2}^{2}=nn'/r+\Lambda_{3}$,
$T_{3}^{3}=n'^2+nn''+\Lambda_{3}$ from (\ref{einsteincomb3}).

All these solutions extend the black hole on the brane into the
bulk. In the BTZ case if we calculate the curvature invariants, we find 
no $r=0$ curvature singularity for the BTZ string-like solution as
expected.
The warp function $f^{2}(\rho)$ gives the shape of the horizon of
the string-like solution while the size of the horizon is defined
by the scale $\sqrt{\alpha}$. 

From our solutions we can notice that consistency of the five-dimensional 
bulk equations requires a fine-tuned relation between the GB coupling 
constant and the AdS$_5$ length (through the five-dimensional cosmological
constant). The use of this fine-tuning gives to
the non-singular horizon the shape of a throat up to the boundary
of the AdS space.

\section{Localization of the Black Hole on the Brane}

In order to complete our solution with the introduction of the
brane we must  solve the corresponding junction conditions given
by the ``boundary'' Einstein equations (\ref{einsteincomb3}) using
the induced metric on the brane given by (\ref{3dmetric}). For the
case when $n(r)$ corresponds to the BTZ black hole (\ref{btz}),
and the brane cosmological constant is given by
$\Lambda_{(3)}=-1/L_3^{2}$, we found that the energy-momentum tensor
is null. Therefore, the BTZ black hole is localized on the brane
in vacuum.

When $n(r)$ is of the form given in (\ref{btzc}), we found
the following traceless energy-momentum tensor 
\begin{equation} 
T_\alpha ^\beta =  \hbox{diag } \left(
\frac{\zeta}{2r^3},\frac{\zeta}{2r^3},-\frac{\zeta}{r^3} \right)\,,
\label{braneEnerMom} 
\end{equation} 
which is conserved on the brane, $\bigtriangledown_\beta T_\alpha ^\beta =0$. 
This property has been studied in \cite{Kofinas:2005a} for codimension-2 
braneworlds in six dimensions; however, it seems to be a property independent
of the bulk dimension for Einstein-Gauss-Bonnet codimension-2
models having an axially symmetric metric like (\ref{5dmetric})
with $a^2(x,\rho)=1$, and obeying junction conditions analogous to
(\ref{einsteincomb3}). If we consider the energy-momentum tensor in 
\cite{Zanelli1996} necessary to sustain the solution (\ref{btzc}),
and we take the limit  $r/L_3<<1$, we get the unexpected
result that it reduces to (\ref{braneEnerMom}) which is necessary
to localize the black hole on the conical 2-brane. A way to understand this
result is that because in this limit $r$ is very small, the black
hole will be localized around the conical singularity and
therefore, any matter will take a distributional form around this
singularity.
Note also that this solution is a result of the presence
of the GB term in the bulk. If we switch off the
GB coupling, then from relations (\ref{173}) and
(\ref{183}) it can be seen that only the BTZ black hole is a
solution.

\section{Conclusions}

We discussed black hole localization on an infinitely thin 2-brane
of codimension-2 and its extension into a five-dimensional AdS
bulk. To have a three-dimensional gravity on the brane we
introduced a five-dimensional Gauss-Bonnet term in the bulk and an
induced gravity term on the 2-brane. We showed that the (2+1) BTZ
black hole and its short-distance-corrected extension can be localized on
the 2-brane, while in the bulk these solutions describe BTZ
strings. Consistency of the five-dimensional bulk equations
requires a fine-tuned relation between the Gauss-Bonnet coupling
constant and the length of the five-dimensional AdS space. The use
of this fine-tuning gives to the non-singular horizon the shape of
a throat up to the boundary of the AdS space.

We did not allow more severe singularities than conical. This
assumption has fixed the deficit angle to a constant value.

This analysis has also been extended to a conical 3-brane embedded in a
six-dimensional bulk as reported in \cite{cptz2}. This study also showed
the relevance of the GB term as it dictates the kind of brane and bulk
matter necessary to sustain a black hole on the brane.

\ack{This work was supported by the NTUA research
program PEVE07. The work of B.C-M. is supported by Fondo Nacional
de Desarrollo Cient\'{i}fico y Tecnol\'ogico (FONDECYT), Chile,
under grant 3070009.}


\section*{References}
\begin{thebibliography}{9}
\bibitem{Chen:2000at}
  Chen J W, Luty M A and Ponton E, 2000, 
  {\it JHEP} {\bf 0009}, 012.
\bibitem{6d}
  Carroll S M, Guica M M, {\it Preprint} arXiv:hep-th/0302067;
  Aghababaie Y {\it et al.}, 2004, 
  {\it Nucl.\ Phys.} B {\bf 680}, 389;
  Nilles H P, Papazoglou A, Tasinato G, 2004, 
  {\it Nucl.\ Phys.} B {\bf 677}, 405;
  Kehagias A, 2004, 
  {\it Phys.\ Lett.} B {\bf 600}, 133.
\bibitem{Cline}
  Cline J M, Descheneau J, Giovannini M, Vinet J, 2003, 
  {\it JHEP} {\bf 0306}, 048.
\bibitem{Kanno:2004nr}
  Kanno S, Soda J, 2004, 
  {\it JCAP} {\bf 0407}, 002.
  Vinet J, Cline J M, 2004, 
  {\it Phys.\ Rev.} D {\bf 70}, 083514;
  Navarro I, Santiago J, 2005, 
  {\it JHEP} {\bf 0502}, 007.
\bibitem{Bostock:2003cv}
  Bostock P, Gregory R, Navarro I and Santiago J, 2004, 
  {\it Phys.\ Rev.\ Lett.}  {\bf 92}, 221601.
\bibitem{Papantonopoulos:2005ma}
  Papantonopoulos E, Papazoglou A, 2005, 
  {\it JCAP} {\bf 0507}, 004.
\bibitem{Kofinas:2005py}
  Kofinas G, 2006, 
  {\it Phys.\ Lett.}  B {\bf 633}, 141.
  Papantonopoulos E, Papazoglou A, 2005, 
  {\it JHEP} {\bf 0509}, 012.
\bibitem{regular}
  Carter B, Battye R A and Uzan J P, 2003, 
  {\it Commun.\ Math.\ Phys.}  {\bf 235}, 289;
  Kolanovic M, Porrati M and Rombouts J W, 2003, 
  {\it Phys.\ Rev.}  D {\bf 68}, 064018;
  Peloso M, Sorbo L and Tasinato G, 2006, 
  {\it Phys.\ Rev.}  D {\bf 73}, 104025;
  Papantonopoulos E, Papazoglou A and Zamarias V, 2007, 
  {\it JHEP} {\bf 0703}, 002.
  Himmetoglu B and  Peloso M, 2007, 
  {\it Nucl.\ Phys.}  B {\bf 773}, 84;
  Kobayashi T and  Minamitsuji M, 2007,
  {\it Phys.\ Rev.}  D {\bf 75}, 104013;
  Burgess C P, Hoover D and  Tasinato G, 2007,
  {\it JHEP} {\bf 0709}, 124;
  Kobayashi T and  Takamizu Y I, 2007,
  {\it Class.\ Quant.\ Grav.} {\bf 25}, 015007;
  Copeland E J and Seto O, 2007, 
  {\it JHEP} {\bf 0708}, 001.
\bibitem{Papantonopoulos:2007fk}
  Papantonopoulos E, Papazoglou A and Zamarias V, 2008, 
  {\it Nucl.\ Phys.} B {\bf 797}, 520-536;
  Minamitsuji M and Langlois D, 2007, 
  {\it Phys.\ Rev.}  D {\bf 76}, 084031.
\bibitem{Chamblin:1999by}
  Chamblin A, Hawking S W, Reall H S, 2000, 
  {\it Phys.\ Rev.} D {\bf 61}, 065007.
\bibitem{glf} 
  Gregory R and Laflamme R, 1993, 
  {\it Phys. Rev. Lett.} {\bf 70}, 2837.
\bibitem{SMS}
  Shiromizu T, Maeda K I and Sasaki M, 2000, 
  {\it Phys.\ Rev.} D {\bf 62}, 024012.
\bibitem{BBH}
  Dadhich N, Maartens R, Papadopoulos P and Rezania V, 2000, 
  {\it Phys.\ Lett.} B {\bf 487}, 1;
  Bruni M, Germani C and Maartens R, 2001, 
  {\it Phys.\ Rev.\ Lett.}  {\bf 87}, 231302;
  Casadio R, Fabbri A and Mazzacurati L, 2002, 
  {\it Phys.\ Rev.} D {\bf 65}, 084040;
  Kofinas G, Papantonopoulos E and  Pappa I, 2002, 
  {\it Phys.\ Rev.} D {\bf 66}, 104014;
  Kofinas G, Papantonopoulos E and Zamarias V, 2002, 
  {\it Phys.\ Rev.}  D {\bf 66}, 104028.
\bibitem{ACMPM}
  Abdalla E, Cuadros-Melgar B, Pavan A B and Molina C, 2006, 
  {\it Nucl.\ Phys.} B {\bf 752}, 40.
\bibitem{EHM}
  Emparan R, Horowitz G T and Myers R C, 2000, 
  {\it JHEP} {\bf 0001}, 007; 
  Emparan R, Horowitz G T and Myers R C, 2000, 
  {\it JHEP} {\bf 0001}, 021.
\bibitem{Kinnersley:zw}
  Kinnersley W and Walker M, 1970, 
  {\it Phys.\ Rev.} D {\bf 2}, 1359.
\bibitem{Banados:1992wn}
  Ba\~nados M, Teitelboim C and Zanelli J, 1992, 
  {\it Phys.\ Rev.\ Lett.}  {\bf 69}, 1849.
\bibitem{Kaloper:2006ek}
  Kaloper N and Kiley D, 2006, 
  {\it JHEP} {\bf 0603}, 077.
  Kiley D, 2007, 
  {\it Phys.\ Rev.} D {\bf 76}, 126002.
\bibitem{Aryal:1986sz}
  Aryal M, Ford L H and Vilenkin A, 1986, 
  {\it Phys.\ Rev.} D {\bf 34}, 2263;
  Achucarro A, Gregory R and Kuijken K, 1995, 
  {\it Phys.\ Rev.} D {\bf 52}, 5729.
\bibitem{CuadrosMelgar:2007jx}
  Cuadros-Melgar B, Papantonopoulos E, Tsoukalas M and Zamarias V, 2008, 
  {\it Phys.\ Rev.\ Lett.}  {\bf 100}, 221601.
\bibitem{Zanelli1996}
  Martinez C and Zanelli J, 1996, 
  {\it Phys.\ Rev.} D {\bf 54}, 3830.
\bibitem{Kofinas:2005a}
  Kofinas G, 2005, 
  {\it Class.\ Quant.\ Grav.} {\bf 22}, L47.
\bibitem{cptz2}
  Cuadros-Melgar B, Papantonopoulos E, Tsoukalas M and Zamarias V, 2009, 
  {\it Nucl.\ Phys.} B {\bf 810}, 246-265.
\end{thebibliography}
\end{document}